\documentclass[11pt,a4paper]{article}
\usepackage{jheppub}
\usepackage{bbm}
\usepackage{pdfpages}
\usepackage{stackrel}
\usepackage{epstopdf}

\input pix.sty

\newcommand{\sumint}[1]{\mbox{$\sum$}\!\!\!\!\!\!\!\int_{#1}}

\renewcommand{\nr}[1]{(\ref{#1})}

\newcommand{\dd}{\mathrm{d}}
\newcommand{\tinymsbar}{{\overline{\mbox{\tiny\rm{MS}}}}}

\newcommand{\gB}{g_\rmii{B}}

\def\lsi{\raise0.3ex\hbox{$<$\kern-0.75em\raise-1.1ex\hbox{$\sim$}}}
\def\gsi{\raise0.3ex\hbox{$>$\kern-0.75em\raise-1.1ex\hbox{$\sim$}}}

\newcommand{\rmii}[1]{{\mbox{\tiny\rm{#1}}}}

\newcommand{\im}{\mathop{\mbox{Im}}}

\newcommand{\Tint}[1]{{\hbox{$\sum$}\!\!\!\!\!\!\!\int\,}_{\!\!\!\!\raise-0.9ex\hbox{$\scriptstyle{#1}$}}}
\newcommand{\Tinti}[1]{{{\Sigma}\!\!\!\!\raise0.3ex\hbox{$\int$}_\rmii{${#1}$}}}
\renewcommand{\Tint}[1]{\sumint{#1}}

\newcommand{\bi}{\begin{itemize}}
\newcommand{\ei}{\end{itemize}}

\newcommand{\hide}[1]{ }

\def\CN{{\cal N}}

\title{Energy momentum tensor correlators in hot Yang-Mills theory: holography confronts lattice and perturbation theory}

\author[1]{K. Kajantie,}
\author[2]{Martin Kr\v{s}\v{s}\'ak}
\author[2]{and Aleksi Vuorinen}

\affiliation[1]{Department of Physics, P.O.Box 64, FI-00014 University of Helsinki, Finland}
\affiliation[2]{Faculty of Physics, University of Bielefeld, D-33501 Bielefeld, Germany}

\emailAdd{keijo.kajantie@helsinki.fi}
\emailAdd{krssak@physik.uni-bielefeld.de}
\emailAdd{vuorinen@physik.uni-bielefeld.de}

\preprint{BI-TP 2012/53, HIP-2013-03/TH}

\abstract{We investigate the behavior of energy momentum tensor correlators in strongly coupled large-$N_c$ Yang-Mills theory at nonzero temperature, working within the Improved Holographic QCD model. In particular, we determine the spectral functions and corresponding imaginary time correlators in the bulk and shear channels, and compare the results to recent perturbative and lattice calculations where available. For the bulk channel imaginary time correlator, for which all three results exist, lattice data is seen to favor the holographic prediction over the perturbative one over a wide range of temperatures.}

\keywords{Holography and Quark Gluon Plasma, AdS/CFT Correspondence}

\begin{document}

\maketitle

\section{Introduction}\label{intro}

Alongside with the realization that the quark gluon plasma (QGP) created at RHIC should be described as a strongly coupled liquid rather than a gas of weakly interacting quasiparticles \cite{Tannenbaum}, holographic methods have become a standard tool in making qualitative --- and in a few cases even quantitative --- predictions for heavy ion physics \cite{CasalderreySolana:2011us}. Perhaps the best known example of this is the famous conjecture of a lower limit for the shear viscosity to entropy ratio, $\eta/s\geq 1/(4\pi)$ \cite{Kovtun:2004de}, which the QGP appears to almost saturate (see also \cite{Romatschke:2009im,Kovtun:2011np,Rebhan:2011vd}). In addition, holography has been used to address many complicated dynamical problems out of the reach of conventional field theory methods, such as strong coupling thermalization and particle production (see e.g.~\cite{Chesler:2010bi,Steineder:2012si} and references therein). Within thermal equilibrium, it has finally been observed that the behavior of many bulk 
thermodynamic quantities near the critical temperature of the deconfinement transition, measurable with lattice methods, can be reproduced to a very good accuracy using holographic models with broken supersymmetry and conformal invariance \cite{kiri2,kiri3,Alanen:2009xs}.

On the field theory side, a major obstacle in the quantitative description of the thermalizing plasma is the inapplicability of lattice methods to real time physics --- or even to the determination of transport coefficients. In the latter case, discussed extensively e.g.~in \cite{Meyer:2011gj}, some progress has recently been made in combining lattice measurements of Euclidean correlators with perturbative results for the corresponding spectral functions \cite{Burnier:2011jq,Burnier:2012ts}. Despite this, important hydrodynamic parameters such as the shear and bulk viscosities are still outside the realm of accurate first principles calculations. Recalling that these quantities are readily available in the strongly coupled limit of a class of large-$N_c$ field theories via the AdS/CFT conjecture, it is not surprising that quite some attention has lately turned towards a quantitative comparison of lattice, perturbative and gauge/gravity predictions for various (mostly Euclidean) correlation functions. Recent 
lattice studies of energy momentum tensor correlators include at least \cite{Huebner:2008as,Iqbal:2009xz,Meyer:2010ii,meyershear}, while related perturbative work has been performed in
\cite{Laine:2010fe,Laine:2010tc,Laine:2011xm,Schroder:2011ht,Zhu:2012be} and holographic calculations in
\cite{teaney,Gubser:2008sz,Gursoy:2009kk,Kajantie:2010nx,Kajantie:2011nx,Springer:2010mf,Springer:2010mw}. Finally, closely related studies of sum rules that the associated spectral functions must obey can be found e.g.~from \cite{Romatschke:2009ng,Meyer:2010gu}, while an analytic study of the UV limit of different correlators was performed in \cite{CaronHuot:2009ns}.

A particularly interesting comparison of lattice, weak coupling and holographic correlators was reported in \cite{Iqbal:2009xz}. There, it was found that lattice data for various Euclidean Green's functions just above the deconfinement temperature of SU(3) Yang-Mills plasma is better described by infinitely strongly coupled ${\mathcal N}=4$ Super Yang-Mills (SYM) theory than by a leading order perturbative calculation in the original theory. Even though there are indications that the inclusion of further perturbative orders acts in the direction of closing the gap between the weak coupling and lattice results \cite{Laine:2010tc,Laine:2011xm}, it is equally worthwhile to attempt to improve the description of the system on the strong coupling side. To this end, in \cite{Kajantie:2011nx} we addressed the determination of the shear channel spectral function and the corresponding imaginary time and coordinate space correlators in the so-called Improved Holographic QCD (IHQCD) model, which exhibits a dynamical 
dilaton field that has the effect of breaking conformal invariance and supersymmetry \cite{kiri2,kiri3}. The study revealed important quantitative effects originating from the loss of conformal invariance near $T_c$, and in addition highlighted the importance of performing similar computations in the more complicated bulk channel, where conformal theories (such as ${\mathcal N}=4$ SYM) lead to vanishing correlation functions.

In the paper at hand, our aim is to continue and extend the treatment of \cite{Kajantie:2011nx} by performing a detailed analysis of the bulk channel of the IHQCD model, concentrating in particular on the spectral function at vanishing external three-momentum and the associated imaginary time correlator. The latter quantity is of special interest to us due to the recent emergence of the corresponding perturbative and lattice results \cite{Meyer:2010ii,Laine:2011xm}, to which we can compare our holographic predictions. In addition to this, we will briefly revisit the shear channel, where a Next-to-Leading Order (NLO) result for the spectral function has been determined since the appearance of \cite{Kajantie:2011nx}, motivating a reanalysis of the IHQCD calculation.

Our paper is organized as follows. In section 2, we review the setup both on the field theory and gravity sides, recalling the most important aspects of the IHQCD model. In section 3, we next write down the fluctuation equations, from which both the shear and bulk spectral functions are determined, and in addition explain the most important steps of the holographic calculation. Section 4 then reviews the existing perturbative results for the quantities of our interest, while sections 5-6 contain our holographic results in the two channels. A more comprehensive discussion of the results is finally left to section 7, where we also draw our conclusions.

Our notation follows closely that explained in section 1 of \cite{Kajantie:2011nx}. In particular, with the exception of the introductory section 2, we will in the following set the AdS radius ${\mathcal L}=1$.

\section{Setup}\label{setup}

\subsection{Field theory}

We work within pure SU($N_c$) Yang-Mills theory at a nonzero temperature $T$, defined by the Euclidean Lagrangian
\ba
 S_\mathrm{E} &=& \int_{0}^{\beta} \! \dd \tau \int \! {\rm d}^{3}\vec{x}
 \, \frac{1}{4} F^a_{\mu\nu} F^a_{\mu\nu} \,,\quad F^a_{\mu\nu} \,\equiv\,  \partial_\mu A^a_\nu - \partial_\nu A^a_\mu + \gB f^{abc} A^b_\mu A^c_\nu\, ,
\ea 	
with $\beta\equiv 1/T$. The energy momentum tensor of the theory takes the form
\begin{equation}\la{eq:T}
 T_{\mu\nu}(x) = \frac{1}{4} \delta_{\mu\nu} F^a_{\alpha\beta} F^a_{\alpha\beta} -F^a_{\mu\alpha} F^a_{\nu\alpha} =\theta_{\mu\nu}(x)+\frac{1}{4}\delta_{\mu\nu}\theta(x)\, ,
\end{equation}
where we have in the latter stage separated the traceless part $\theta_{\mu\nu}$ and the anomalous trace
\begin{equation}
\theta(x)\equiv T_{\mu\mu}=\frac{\beta(g)}{2g}F_{\mu\nu}^aF_{\mu\nu}^a\, ,
\end{equation}
in which $\beta(g)$ denotes the beta function of the theory.

In this paper, we are interested in correlation functions of the shear and bulk operators $T_{12}$ and  $\theta$ of the above theory, of which the latter we can furthermore replace by the simpler quantity $T_{ii}$, as correlators of $T_{00}$ are known to reduce to contact terms (see e.g.~\cite{Meyer:2010ii}). This implies that the retarded correlators we study obtain the forms
\ba
G_s^R(\omega,\mathbf{k}=0)&=&-i\int\! {\rm d}^4x\,  e^{i \omega t} \theta (t) \langle[
T_{12}(t,\vec{x}),T_{12}(0,0) ]\rangle \label{Gsdef}\, \\
G_b^R(\omega,\mathbf{k}=0)&=&-i\int\! {\rm d}^4x\,  e^{i \omega t} \theta (t) \langle[\frac{1}{3}
T_{ii}(t,\vec{x}),\frac{1}{3}T_{jj}(0,0) ]\rangle \label{Gbdef} \, ,
\ea
while the corresponding (zero three-momentum) spectral functions read
\ba
\rho_{s,b} (\omega,T)&=&\im G_{s,b}^R(\omega,\mathbf{k}=0)\, .
\ea
The relation between these functions and the associated transport coefficients (the shear and bulk viscosities) is finally given by
\ba
\eta&=&\lim_{\omega\to 0}\frac{\rho_s(\omega,T)}{\omega}\, ,\label{eta} \\
\zeta&=&\lim_{\omega\to 0}\frac{\rho_b(\omega,T)}{\omega}\, .\label{zeta}
\ea
It is also good to recall that in conformal theories, in which the beta function is zero, we have $T_{\mu\mu}=0$, and thus vanishing bulk correlators and viscosity.

\subsection{Dual gravity system} \label{dual}

In the IHQCD model of \cite{kiri2,kiri3}, the gravitational system involves a gravity+dilaton type action
\begin{equation}
S=\frac{1}{16 \pi G_5}\int {\rm d}^5x\sqrt{-g} \left[R -\frac{4}{3}(\partial\phi)^2 +V(\phi) \right]\, ,  \label{action}
\end{equation}
while the background metric takes the generic form
\begin{equation}
ds^2=b^2(z)\left[-f(z)dt^2+d{\bf x}^2 +{dz^2\over f(z)}\right]\, ,  \label{metric}
\end{equation}
where the radial coordinate $z$ is chosen so that the boundary is located at $z=0$. The functions $\phi(z)$, $f(z)$ and $b(z)$ appearing here are determined from the Einstein equations
\begin{eqnarray}
\dot{W}&=& 4 b W^2 -\frac{1}{f}(W\dot{f} +\frac{1}{3}b V ),  \label{Einstein1}\\
\dot{b} &=&-b^2 W\,, \label{Einstein2}\\
\dot{\lambda} &=&\frac{3}{2}\lambda \sqrt{b \dot{W}}\,,\label{Einstein3}\\
\ddot{f} &=& 3\dot{f}b W\,,\label{Einstein4}
\end{eqnarray}
in which the dot denotes a derivative with respect to $z$, and we have defined $\lambda (z)= e^{\phi(z)}$. As the notation suggests, this function is found to be dual to the 't Hooft coupling on the field theory side, $\lambda_c\equiv g^2 N_c$, while
\begin{equation}
\beta(\lambda)\equiv\frac{\dot{\lambda}}{\dot{b}/b}
\la{betafn}
\end{equation}
is related to the field theory beta function.

In the vicinity of the boundary, the metric function $b(z)$ is required to satisfy
\begin{equation}
b(z)\stackrel[z\to 0]{}{\rightarrow} \frac{\mathcal{L}}{z}\, ,
\end{equation}
where $\mathcal{L}$ is the curvature radius of AdS space. For the other functions, the UV limits are obtained from the running of the field theory coupling, while the IR behaviors are determined by requiring that the model satisfy the confinement criterion of a linear glueball spectrum $m^2\sim \mathrm{integer}$ \cite{kiri2} (see also appendices A and B of \cite{Kajantie:2011nx}). The dilaton potential we use follows the choice of \cite{Jarvinen:2011qe}, reading
\begin{equation}
V(\lambda)=\frac{12}{\mathcal{L}^2}\left[
1+\frac{88}{27}\lambda + \frac{4619}{729}\lambda^2\frac{ \sqrt{1+\ln(1+\lambda)}}{(1+\lambda)^{2/3}}
\right]\, . \label{V}
\end{equation}
It is constructed to reproduce the small $\lambda$ expansion
\begin{equation}
V(\lambda)=\frac{12}{\mathcal{L}^2}\left[
1+\frac{88}{27}\lambda + \frac{4619}{729}\lambda^2+{\mathcal O}(\lambda^3)
\right]\, ,
\end{equation}
where the coefficients in front of $\lambda$ and $\lambda^2$ are determined by matching the holographic beta function of eq.~\nr{betafn} to the perturbative 2-loop result with the identification $\lambda=\lambda_c/(8\pi^2)$. The $\lambda^2$ term in eq.~(\ref{V}) is finally multiplied by the factor
\begin{equation}
\frac{\sqrt{1+\ln(1+\lambda)}}{(1+\lambda)^{2/3}}
\end{equation}
to ensure that the model satisfies the confinement criterion.

With the gravity system specified, we can now use eqs.~(\ref{Einstein1})-(\ref{Einstein4}) to solve for those bulk field configurations that exhibit a horizon at $z=z_h$ (i.e.~satisfy $f(z_h)=0$) and thus correspond to an equilibrium state of the field theory. This enables us to determine a host of equilibrium thermodynamical quantities such as the pressure of the system \cite{kiri3,Alanen:2011hh}, which we match to its leading order counterpart in perturbative large-$N_c$ Yang-Mills theory. This leads to one further matching relation
\begin{equation}
\frac{\mathcal{L}^3}{4\pi G_5}=\frac{4 N_c^2}{45 \pi^2}\, ,
\la{ellcube}
\end{equation}
which fixes the last unknown parameter of the model.

\section{The calculations}\label{calculations}

To determine the correlation functions of the field theory operators $T_{12}$ and $T_{ii}$ using holography, we follow the steps laid out in \cite{Gubser:2008sz}. First, we introduce perturbations around the background metric of eq.~(\ref{metric}),
\ba
g_{00} &=& b^2 f \left(1+\epsilon H_{00}\right)\,, \;\;
g_{11} = b^2 \left(1+\epsilon H_{11}\right)\, , \;\;
g_{12}=\epsilon b^2 H_{12}\, , \;\;
g_{55} =\frac{b^2}{f} \left(1+\epsilon H_{55}\right)\,,\;\;\; \label{bulkpert}
\ea
where $\epsilon$ is a power counting parameter. With these definitions, the  perturbation $H_{11}$ becomes dual to the operator $\frac{1}{3}T_{ii}$, while $H_{12}$ corresponds to the shear operator $T_{12}$.  Expanding the Einstein equations to first order in $\epsilon$, we then find that the metric fluctuations must satisfy the equations
\ba
&&\ddot H_{12}+ {d\over dz}\log(b^3f)\dot H_{12}+
{\omega^2\over f^2} H_{12}=0 \,, \label{fluctshear} \\
&&\ddot H_{11}+ {d\over dz}\log(b^3fX^2)\dot H_{11}+
\biggl({\omega^2\over f^2} -{\dot f\,\dot X\over fX}\biggr)H_{11}=0\,,\label{fluctbulk}
\ea
where we have set the corresponding three-momentum $\mathbf{k}$ to zero and defined
\ba
X(\lambda) &\equiv& \frac{\beta (\lambda)}{3 \lambda}. \label{Xdef}
\ea
These equations are to be solved using purely infalling boundary conditions at the horizon, most conveniently implemented via an analytic expansion around $z= z_h$,
\begin{equation}
H_{12/11}(z\rightarrow z_h)=(z-z_h)^{i\omega/\dot{f}_h}[1 +d_1(z-z_h)+ d_2(z-z_h)^2 +\dots ]\, . \label{Hexp}
\end{equation}
The equations of motion (\ref{fluctshear})-(\ref{fluctbulk}) are then straightforwardly solvable using Mathematica.

With the metric fluctuations at hand, a standard recipe provides us with rather simple forms for the shear and bulk spectral functions,
\ba
\rho_s(\omega,T)&=&\frac{f(z) b(z)^3}{16 \pi G_5}\, \frac{\im \dot{H}_{12}(z)H^*_{12}(z)}{|H_{12}(z\rightarrow 0)|^2}\label{rhos}\, ,\\
\rho_b(\omega,T)&=&\frac{6X(z)^2\,f(z) b(z)^3}{16 \pi G_5} \,  \frac{\im \dot{H}_{11}(z)H^*_{11}(z)}{|H_{11}(z\rightarrow 0)|^2}\, ,\label{rhob}
\ea
in which the factors $|H_{12/11}(z\rightarrow 0)|^{-2}$ account for the fact that the functions in eq.~(\ref{Hexp}) have not been normalized to unity at the boundary. It can furthermore be shown that the expressions (\ref{rhos}) and (\ref{rhob}) are in fact independent of $z$, i.e.~can be evaluated at any value of the radial coordinate (cf.~\cite{Kajantie:2011nx} for a more detailed discussion of this issue). For practical reasons, we choose to do so infinitesimally close to the horizon, where a use of eq.~(\ref{Hexp}) as well as the identity $f(z\to z_h) = \dot{f}(z_h)(z-z_h)+{\mathcal O}((z-z_h)^2)$ leads us to the expressions
\ba
\rho_s(\omega,T)&=&\frac{s(T)}{4 \pi}\frac{\omega}{|H_{12}(z\rightarrow 0)|^2}\,, \label{rhos1} \\
\rho_b(\omega,T)&=&6X_h^2 \frac{s(T)}{4 \pi}\frac{\omega}{|H_{11}(z\rightarrow 0)|^2}\, ,\label{rhob1}
\ea
in which $X_h\equiv X(z_h)$, and $s(T)=b_h^3/(4G_5)$ denotes the entropy. One should note that these results mix the IR and UV scales of the system in a way that will be seen to result in very interesting large-$\omega$ behavior of the spectral functions in the following sections.

\section{Perturbative limit}\label{results}

Before proceeding to the results of our holographic calculations, let us briefly review what is known about the behavior of the shear and bulk spectral functions in weakly coupled SU($N_c$) Yang-Mills theory. This is helpful in particular for the analysis of the UV (large-$\omega$) behavior of our results, as due to asymptotic freedom all physical correlators are expected to reduce to their perturbative limits as $\omega\to\infty$.

At the moment, the $\mathbf{k}=0$ spectral functions of both channels are known up to and including their respective NLO terms in perturbation theory \cite{Laine:2011xm,Zhu:2012be}. In the shear case, we can read off the result from eq.~(4.1) of \cite{Zhu:2012be}, obtaining (note an additional factor of -1/16 due to differing definitions of the shear operator)
\ba
\frac{\rho_s(\omega,T)}{d_A}&=&\frac{\omega^4}{160\pi}\bigl( 1 + 2 n_{\frac{\omega}{2}} \bigr)\Bigg\{1-\frac{10\lambda_c}{16\pi^2}\bigg(\frac{2}{9}+\phi_T^\eta({\omega\over T})\bigg)\Bigg\}+
{\mathcal O}(\lambda_c^2)\label{rhosper0} \\
&\stackrel[\omega\to \infty]{}{\rightarrow}& \frac{1}{160\pi}\,\omega^4 \, , \label{rhosper1}
\ea
where $d_A\equiv N_c^2-1$, $n_x\equiv 1/(e^{x/T}-1)$, and $\phi_T^\eta(\omega/T)$ is a numerically evaluatable dimensionless function that behaves like $T^6/\omega^6$ in the $\omega\to\infty$ limit. It should be noted that this result misses a number of terms proportional to $\omega\,\delta(\omega)$, which give important contributions to the shear sum rule but are only known to leading order, cf.~e.g.~eq.~(4) of \cite{meyershear}.

In the bulk channel, the perturbative spectral function consistent with our earlier definitions is obtainable from eq.~(4.1) of \cite{Laine:2011xm}. Multiplying this result by $1/9$ and choosing the constant $c_\theta$ as $g^2 c_\theta = \frac{\beta(\lambda_c)}{4\lambda_c}$, where $\beta(\lambda_c)$ is the beta function of Yang-Mills theory, we obtain
\ba
\frac{\rho_b(\omega,T)}{d_A}&=&\frac{\omega^4}{576\pi}\frac{\beta(\lambda_c)^2}{\lambda_c^2}\bigl( 1 + 2 n_{\frac{\omega}{2}} \bigr)\Bigg\{1+\frac{\lambda_c}{8\pi^2}\bigg(\frac{44}{3}\ln\frac{\bar{\mu}}{\omega}+\frac{73}{3}+8\phi_T^\theta({\omega\over T})\bigg)\Bigg\}\nonumber \\
&+&{\mathcal O}(\lambda_c^4) \label{rhobper0} \\
&\stackrel[\omega\to \infty]{}{\rightarrow}&
\frac{121\omega^4}{324(4\pi)^5}\lambda_c^2  , \label{rhobper1}
\ea
where $\phi_T^\theta(\omega/T)$ is again a numerical function, whose behavior was analyzed in quite some detail in \cite{Laine:2011xm}. An important difference to the shear channel result is clearly the appearance of the 't Hooft coupling in the leading large-$\omega$ behavior of eq.~(\ref{rhobper1}). Together with the realization that the renormalization scale, with which the coupling runs, is in the limit $\omega\gg T$ necessarily proportional to $\omega$, this implies that the leading UV behavior of the bulk spectral function takes the form of a $T$-independent constant times $\omega^4/(\ln\,\omega/\Lambda_\tinymsbar)^2$. In addition, one should note that in the bulk channel, no terms of the type $\omega\delta(\omega)$ appear at least at the orders considered above.

When analyzing the perturbative results for both the shear and bulk spectral functions, an important thing to note is that even at high temperatures --- and thus weak coupling --- the above expressions are not valid in the limit of very small $\omega$. This is due to the multitude of soft scales that enter the calculation at small momentum exchange and require complicated resummations to be performed when entering the regions of $\omega$ of order $gT$, $g^2T$ and ultimately $g^4T$ (see e.g.~\cite{Arnold:2003zc}). While with the Hard Thermal Loop resummation performed in \cite{Laine:2011xm} the above bulk result should be correct down to momenta of order $\omega\sim gT$, it is clear from the plots of \cite{Laine:2011xm,Zhu:2012be} that in both the shear and bulk cases, the perturbative results begin to lose accuracy when $\omega\lesssim T$. In particular, this implies that even the leading order transport coefficients are not available from the above expressions, and that when comparing our holographic 
results to them, one should only expect quantitative agreement at $\omega\gg T$.

\section{Holographic results in the shear channel}

\begin{figure}[t]
\centering
\includegraphics[width=0.48\textwidth]{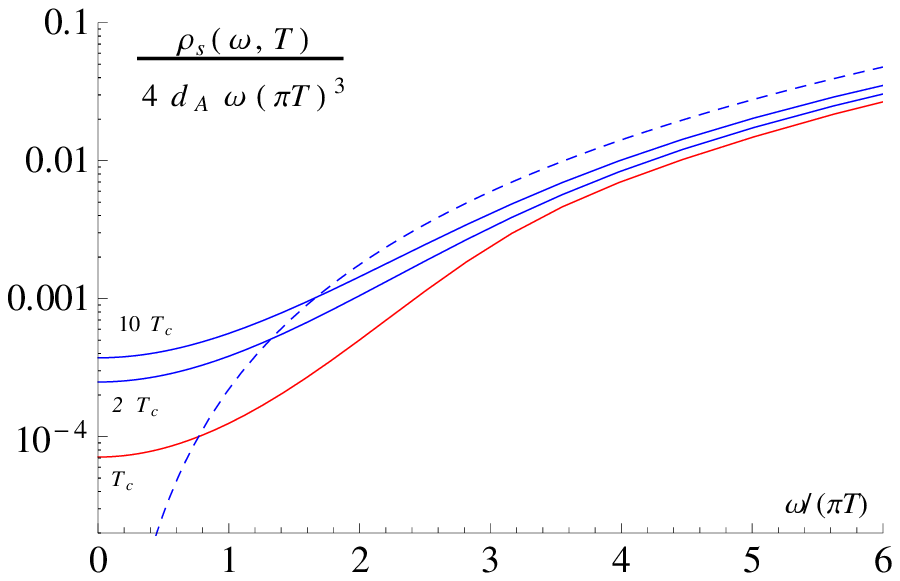}$\;\;\;$ \includegraphics[width=0.48\textwidth]{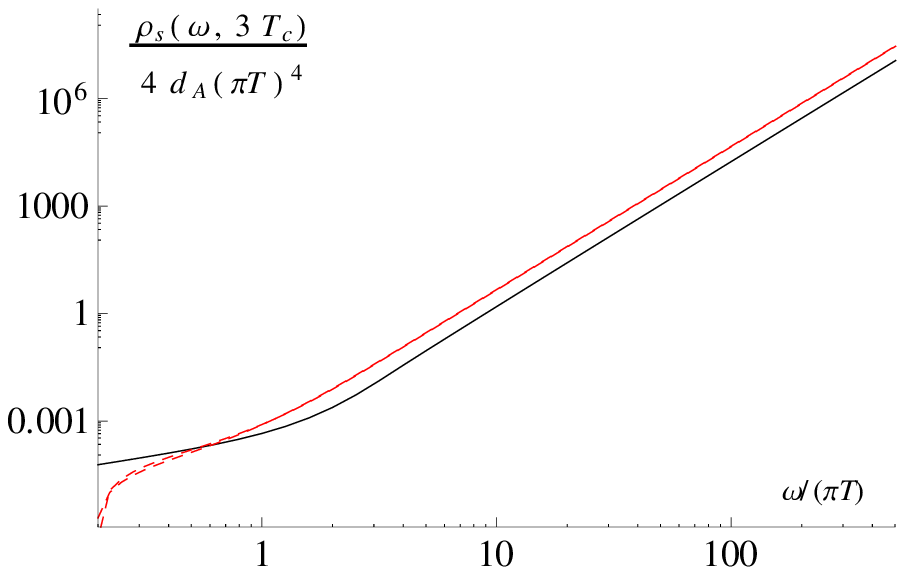}
\caption{\small Left: the IHQCD shear channel spectral function displayed in the region of small frequencies for three different temperatures. The dashed curve represents the large-$\omega$ limit of the SYM result, cf.~\cite{Kajantie:2010nx}. Right: the behavior of the spectral
function at large frequencies for the case of $T=3T_c$. The black curve stands for our IHQCD result, while the two dashed red lines denote the NLO perturbative result evaluated with two different renormalization scales \cite{Zhu:2012be}.}
\label{figshear}
\end{figure}

The shear spectral function was first determined within IHQCD in \cite{Kajantie:2011nx}, and is reproduced for vanishing external three-momentum in fig.~\ref{figshear}. Comparing to the corresponding result in the conformal $\CN=4$ SYM theory \cite{teaney,Kajantie:2010nx}, whose asymptotic behavior is represented by the dashed blue curve on the left, we see that the effects of conformal invariance breaking are largest near the IHQCD deconfinement temperature $T_c$, while already at $T=10T_c$ the IHQCD result is rather close to the SYM one. In the $\omega\to 0$ limit, each of the curves furthermore reproduces the well known result of $\eta/s = 1/(4\pi)$, equally valid in IHQCD as in the SYM theory.

Proceeding to larger values of $\omega$, we display the behavior of the shear spectral function on a log-log scale in fig.~\ref{figshear} (right), where it is further compared with the perturbative result of eq.~(\ref{rhosper0}). This reveals a clear discrepancy between the two results in the UV region, which one can understand using the analytic WKB calculation of \cite{Kajantie:2011nx}. With the help of eq.~(\ref{ellcube}), one namely easily obtains as the limiting behavior of the IHQCD shear spectral function
\ba
\rho_s(\omega,T)&\stackrel[\omega\to \infty]{}{\rightarrow}& \frac{N_c^2}{360\pi}\,\omega^4 ,\label{rhosas}
\ea
which deviates from the perturbative limit of eq.~(\ref{rhosper1}) by a factor of $4/9$. This should not come as a surprise considering that IHQCD is only a two-derivative model, but nevertheless highlights its limitations in describing the UV dynamics of the physical theory.

\section{Holographic results in the bulk channel}

\begin{figure}[t]
\begin{center}
 \includegraphics[width=0.6\textwidth]{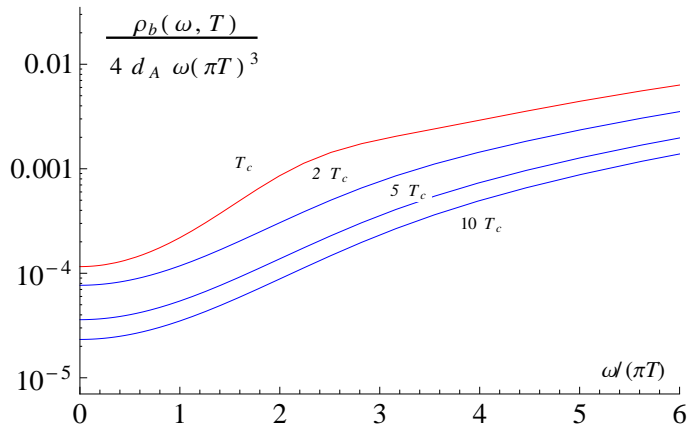}
 \end{center}
 \caption{The bulk spectral function evaluated for a set of different temperatures in the region of small $\omega$.  The temperature dependence of the bulk viscosity can be read off from intercepts of the curves at $\omega=0$.}
 \label{figrhobsmallomega}
 \end{figure}

The bulk channel has been studied in various conformality breaking holographic models already in several works \cite{Gubser:2008sz,Gursoy:2009kk,buchelgursoy}, but typically concentrating only on the behavior of the bulk viscosity to entropy ratio as a function of $T$. In this section, our aim is to extend this treatment to the evaluation of the full IHQCD bulk spectral function.

Beginning from the limit of small frequencies, we first display in fig.~\ref{figrhobsmallomega} the behavior of $\rho_b(\omega,T)$ for temperatures ranging from $T_c$ to $10T_c$ (cf.~also fig.~\ref{figshear} (left)). In accordance with our expectations, we observe a decrease in the values of the function with increasing $T$, signifying the approach of the system towards the conformal limit. From the intercepts of the curves at $\omega=0$, one can furthermore read off the values of the bulk viscosity $\zeta$ at different temperatures, leading to a behavior consistent with that shown in fig.~8 of \cite{Gursoy:2009kk}.

\begin{figure}[t]
\begin{center}
 \includegraphics[width=0.48\textwidth]{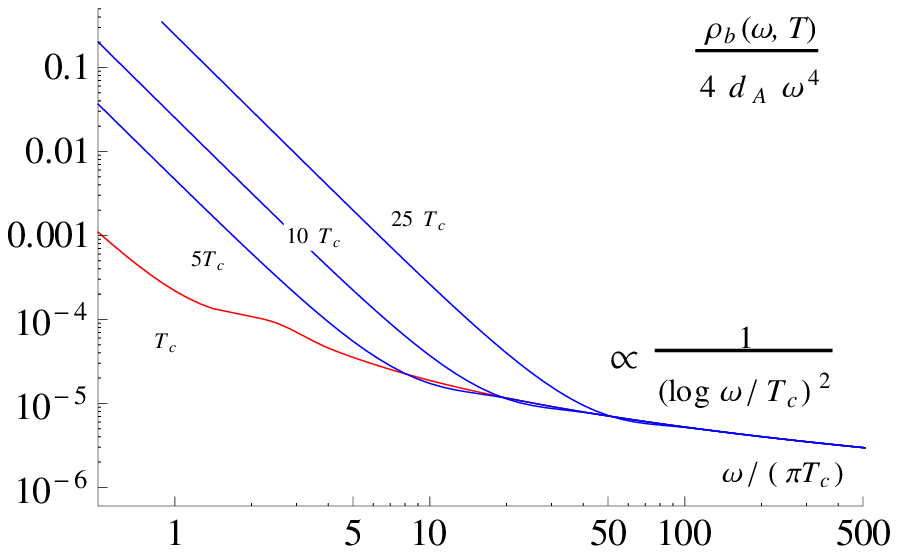}$\;\;\;$ \includegraphics[width=0.48\textwidth]{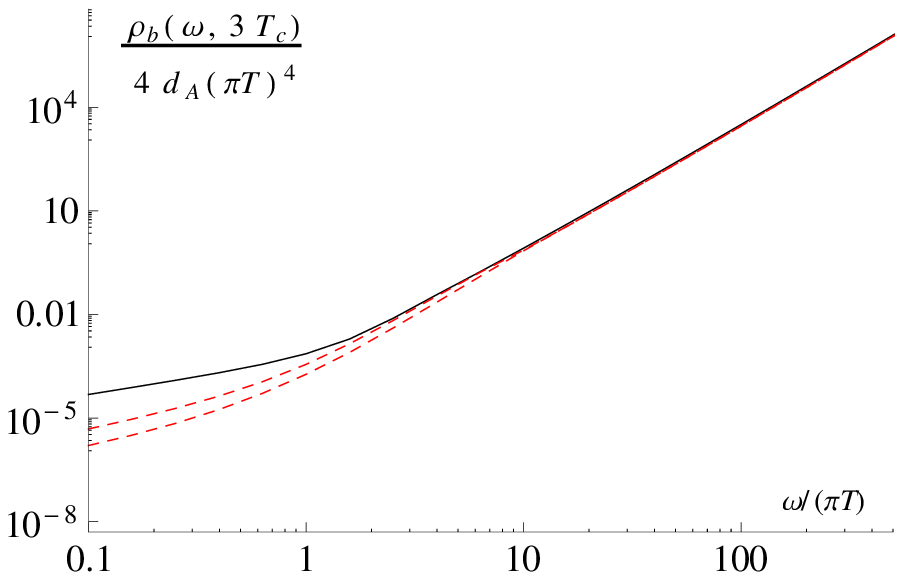}
 \end{center}
 \caption{Left: the behavior of the IHQCD bulk spectral function shown over a wide range of frequencies at three different temperatures. Note that unlike in most of our other plots, $\omega$ has here been scaled dimensionless by $T_c$ and not $T$. Right: a comparison of the IHQCD (solid black curve) and perturbative (red dashed lines) results for the bulk spectral function. The perturbative result is taken from \cite{Laine:2011xm}.}
 \label{figrhoblargeomega}
 \end{figure}

Next, we study the large-$\omega$ behavior of the bulk spectral function in fig.~\ref{figrhoblargeomega}. On the left, we demonstrate, how the combination $\rho_b(\omega,T)/\omega^4$ undergoes a sharp transition from a $T$-dependent $1/\omega^3$ behavior at small frequencies towards a $T$-independent 
$1/(\ln\omega/T_c)^2$ limit at $\omega\gg T$. On the right, we on the other hand specialize to the case of $T=3T_c$, displaying the holographic result together with the perturbative one, eq.~(\ref{rhobper0}). This figure demonstrates a remarkable fact: not only is the form of the asymptotic $1/(\ln\omega/\Lambda_\tinymsbar)^2$ behavior (note that $T_c\sim\Lambda_\tinymsbar$) of the perturbative result reproduced by our IHQCD calculation, but even the overall coefficient in eq.~(\ref{rhobper1}) appears to agree with our numerics. We find this possibly coincidential fact very surprising, considering the missing $4/9$ factor encountered in the shear spectral function. We see no a priori reason, why corrections from higher derivative terms in the holographic action should be present in the shear channel but absent from the bulk one.

Ideally, it would of course be pleasing to be able to derive the logarithmic UV behavior of the bulk spectral function analytically, following a WKB expansion similar to that performed in the shear channel in \cite{Kajantie:2011nx}. In the bulk case, this, however, turns out to be a rather demanding task due to the appearance of logarithmic terms in the fluctuation equation, originating from the $z\to0$ limit of the quantity
\be
3X= {\beta\over\lambda}={d\log\lambda\over d\log b}\stackrel[z\to 0]{}{\rightarrow}{1\over\log z}\, .
\ee
Scaling the radial variable according to $z\to z'=\omega z$, the large-$\omega$ (i.e.~small-$z$) limit of eq.~(\ref{fluctbulk}) namely becomes (with $\Lambda$ denoting an arbitrary scale parameter)
\ba
\ddot H_{11}+\Bigg\{-{3\over z}\biggl(1+{4\over 9(\log \Lambda z/\omega )^2}\biggr)
+{2\over z|\log\Lambda z/\omega |}\Bigg\}\dot H_{11}+H_{11}&=&0 \, ,
\ea
which without the logarithmic terms would lead to the usual $\omega^4$ behavior of the spectral function. In the presence of the conformality breaking $\omega$-dependence, solving the equation however becomes much harder, and in particular leads to the numerically verified appearance of logarithmic suppression in the spectral function. It is in any case worth noting that the resulting $1/(\ln\omega/T_c)^2$ behavior of $\rho_b(\omega,T)$ enters eq.~(\ref{rhob1}) solely through the $z\to0$ limit of $H_{11}(z)$, and not via the factor $X_h^2\sim \beta(\lambda(z_h))^2/\lambda(z_h)^2$. This is in clear contrast with the perturbative limit in eq.~(\ref{rhobper1}), in which the logarithmic behavior is due to the running of the gauge coupling.

\begin{figure}[t]
\begin{center}
  \includegraphics[width=0.48 \textwidth]{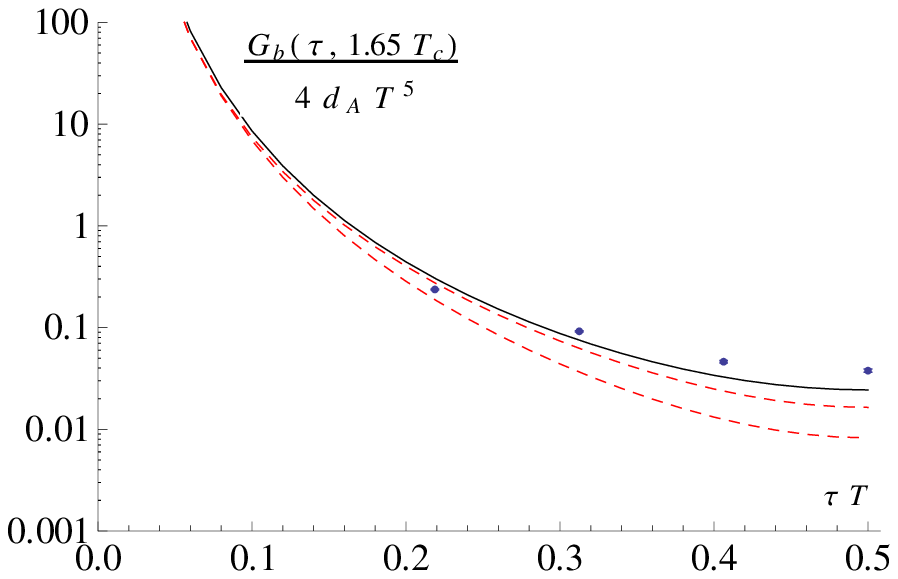}
  $\;\;\;$
   \includegraphics[width=0.48 \textwidth]{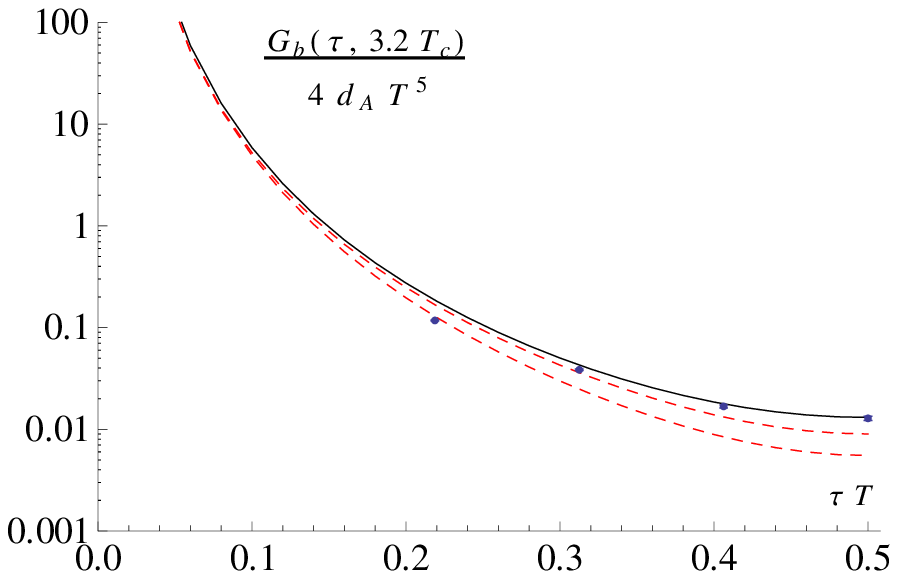}
\end{center}
 \caption{The bulk channel imaginary time correlator of eq.~(\ref{Gtau}) evaluated for two different temperatures in IHQCD (solid black curves) and perturbation theory (red dashed curves), and compared with the lattice data points of \cite{Meyer:2010ii}. The perturbative result is again taken from \cite{Laine:2011xm}.}
 \label{figbim}
 \end{figure}

Having the bulk spectral function now at hand, a natural application is clearly the determination of the corresponding imaginary time correlator, for which both perturbative and lattice results exist. To this end, we plug our function $\rho_b(\omega,T)$ to the relation
\ba
G(\tau,T)&=&\int_0^\infty\frac{d\omega}{\pi}\rho_b(\omega,T)\frac{\cosh\left[\left(
\frac{\beta}{2} -\tau\right) \pi \omega\right]}{\sinh\left( \frac{\beta}{2}\omega\right)}\, ,\quad \beta\equiv 1/T\, ,
\label{Gtau}
\ea
obtaining the result displayed in fig.~\ref{figbim}. Our holographic prediction 
is seen to agree with the lattice data better than the weak coupling result over a wide range of temperatures, the difference being (not surprisingly) most pronounced close to $T_c$. 

\begin{figure}[t]
\begin{center}
 \includegraphics[width=0.6\textwidth]{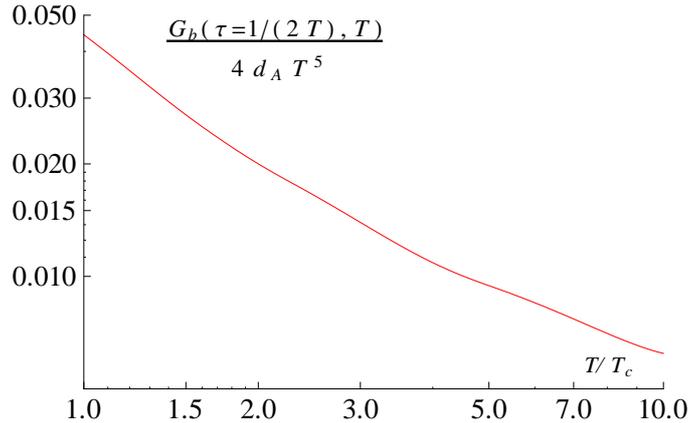}
 \end{center}
 \caption{The imaginary time correlator of eq.~(\ref{Gtau}), normalized by $T^5$ and plotted as a function of temperature at the symmetry point $\tau=1/(2T)$.}
 \label{figbrhosymmpoint}
 \end{figure}

Finally, a different way of inspecting the imaginary time correlator is to look at its value at the symmetry point $\tau=1/(2T)$ as a function of temperature. This we do in fig.~\ref{figbrhosymmpoint}, where $G(\tau=1/(2T),T)$ is displayed, normalized dimensionless by $T^5$. The plot indicates a rapid decrease in the quantity as the temperature is raised above $T_c$, to be contrasted with the slow increase of the corresponding quantity in the shear channel, shown in fig.~6 of \cite{Kajantie:2011nx}. This fact can clearly be attributed to the system approaching conformality in the limit of high temperatures.

\section{Conclusions}\label{conclusions}

In the paper at hand, we have studied finite temperature correlation functions of the energy momentum tensor of large-$N_c$ Yang-Mills theory, concentrating on the shear $\langle T_{12}T_{12}\rangle$ and bulk $\langle T_{ii}T_{jj}\rangle$ channels at vanishing external three-momentum. In particular, after determining the spectral functions and associated imaginary time correlators in the Improved Holographic QCD (IHQCD) model \cite{kiri2,kiri3}, we performed a detailed comparison of our results with state-of-the-art perturbative and lattice works. Clearly, the domains of validity of the three methods do not always overlap. Perturbation theory requires the gauge coupling to be small, which is formally only realized at asymptotically large $T$ or $\omega$, while holographic methods work best in the strongly coupled, yet conformal limit, to which systematic corrections are accounted for in the IHQCD model. Finally, while being a fundamentally nonperturbative first principles method, lattice QCD is unfortunately 
restricted to the Euclidean formulation of the theory, and thus only provides results for a limited set of observables.

Comparing the IHQCD shear and bulk spectral functions with their perturbative counterparts, cf.~figs.~\ref{figshear}-\ref{figrhoblargeomega}, we witnessed an expected pattern, in which conformal invariance breaking effects were seen to be largest near $T_c$, but rapidly decrease with increasing temperature. Furthermore, we saw that in the large-$\omega$ limit of both the shear and bulk channels, the parametric dependence of the perturbative spectral functions on $\omega$ ($\omega^4$ and $\omega^4/(\ln\,\omega/\Lambda_\tinymsbar)^2$, respectively) was correctly reproduced by IHQCD. A closer inspection further revealed that while in the shear channel the $\omega\to\infty$ limit of the perturbative result was larger than the IHQCD one by a factor 9/4, surprisinly in the bulk channel the asymptotic limits perfectly coincide. We find this quite remarkable, considering that the nonzero value of the bulk correlator is entirely due to the conformal invariance breaking built into IHQCD.

For a set of Euclidean quantities --- the imaginary time correlation functions --- we were able to perform comparisons between IHQCD, perturbation theory and lattice Monte Carlo results. A direct comparison of the bulk channel correlator $G_b(\tau,T)$ was performed at two temperatures, $1.65T_c$ and $3.2T_c$. The results showed the lattice data consistently prefering the holographic prediction, though at higher temperatures the difference was seen to somewhat diminish.

Finally, we note that in a recent paper \cite{Gursoy:2012bt}, an IHQCD calculation closely related to ours was performed for the correlators of the pseudoscalar operator $\tr F_{\mu\nu}\tilde{F}_{\mu\nu}$. It is interesting to compare the results reported in section 4 of this paper to ours, concerning in particular the asymptotic large-$\omega$ behavior of the $\mathbf{k}=0$ spectral function. While perturbative arguments suggest that the pseudoscalar spectral function should behave similarly to our bulk result (cf.~ref.~\cite{Laine:2011xm}), the authors of \cite{Gursoy:2012bt} argue that their numerical data at large frequencies is consistent with a pure $\omega^4$ behavior. This being the case, it would clearly be crucial to understand the physical origin of the differing behavior, perhaps by performing a WKB type expansion in the asymptotic region of both channels. This calculation, as well as an analysis of the correlation functions of the $\tr F_{\mu\nu}F_{\mu\nu}$ operator, we however leave for the
future.

\section*{Acknowledgments}
We thank U.~G\"ursoy, E.~Kiritsis, and Yan Zhu for useful discussions, as well as H.~Meyer for providing us his lattice results in a tabulated form. This work was supported by the Sofja Kovalevskaja programme 
of the Alexander von Humboldt foundation, the DFG graduate school \textit{Quantum Fields and Strongly Interacting Matter} as well as the ESF network \textit{Holographic methods for strongly coupled systems (HoloGrav)}.

\end{document}